\documentclass[prl,aps]{revtex4}
\usepackage{graphicx}

\def\n{\langle n \rangle}

\begin{document}
\title{Critical behavior of a bounded Kardar-Parisi-Zhang equation}


\author{Miguel A. Mu\~ noz}
\affiliation{Departamento de Electromagnetismo y F{\'\i}sica de la Materia, 
Universidad de Granada, 
Fuentenueva s/n, 18071 Granada, Spain}

\author{Francisco de los Santos}
\affiliation{Center for Polymer Studies and Department of Physics \\
Boston University, Boston, MA 02215, USA}

\author{Abdelfattah Achahbar}
\affiliation{Departement de Physique, Faculte des Sciences, \\ B.P. 2121
M'hannech, 93002 Tetouan, Morocco}

\date{today}

\begin{abstract}
A host of spatially extended systems, both in physics and in other 
disciplines, are well described at a coarse-grained 
scale by a Langevin equation with multiplicative-noise. 
Such systems may exhibit non-equilibrium phase transitions, which
can be classified into universality classes. 
Here we study in detail one of such classes that 
can be mapped into a Kardar-Parisi-Zhang (KPZ) interface equation 
with a positive (negative) non-linearity in the presence of a bounding lower 
(upper) wall.
The wall limits the possible values taken by the height variable, 
introducing a lower (upper) cut-off, and induce a phase transition 
between a  pinned (active) and a depinned (absorbing) phase.
This transition is studied here using mean field and field theoretical arguments,  
as well as from a numerical point of view.  
Its main properties and critical features, as well as some 
challenging theoretical difficulties, are reported.  
The differences with other multiplicative noise and bounded-KPZ universality 
classes are stressed, and the effects caused by the introduction of 
``attractive'' walls, relevant in some physical contexts, are also analyzed.

\end{abstract}
\maketitle

\section{Introduction}
Non-equilibrium phase transitions occurring in systems amenable to be
described by Langevin equations including a multiplicative noise (MN) term
are the subject of current intense studies. This embraces a broad variety
of systems both in physics and in other disciplines.
A phenomenology much richer and complex than that appearing in equilibrium systems,
including counterintuitive behaviors,
has been reported to appear in these, typically non-equilibrium, situations.
See \cite{Review,Sancho} for detailed introductions to this growing field, 
including many different realizations.

The interest in MN problems is enlarged even further, 
because of the existing mappings between them and
other prototypical non-equilibrium problems \cite{Review}.
A well known instance is the Kardar-Parisi-Zhang (KPZ) equation, 
describing the kinetic roughening transition of generic interfaces under
non-equilibrium conditions \cite{KPZ,Barabasi,HZ}, which
can be mapped onto a MN Langevin equation, by 
performing the so called Cole-Hopf transformation linking the interface 
height at each point with the activity field of the MN equation.
If an interface under consideration is described by 
the KPZ equation and it is physically limited by a wall, {\it i.e.} 
if the heights cannot be larger of smaller than a certain value, 
then this problem, {\it bounded KPZ}, can
be mapped into a multiplicative noise equation by employing  
the abovementioned transformation \cite{MN1,MN2,Review}.
The bounded KPZ equation may experience, as parameters
are varied, a phase transition from a depinned phase in which the
interface escapes with probability one from the wall,
to a pinned phase characterized by a finite expectation value
of the stationary averaged height (measured from the wall).
In the MN language ({\it i.e.} after employing the Cole-Hopf transformation)
the pinning-depinning transition corresponds to a critical point
separating an {\it absorbing phase} in which the order parameter goes
exponentially to zero (depinned phase)
to an {\it active phase} in which the order parameter takes a 
non-vanishing average value.

Surprisingly enough, it was shown a few years ago that the introduction
of ``upper'' or ``lower'' walls into a given KPZ equation 
(with a fixed non-linearity 
sign) lead to quite different phenomenologies. The origin of this can be tracked
down to the fact that the KPZ equation is not invariant upon inverting
the height (see  \cite{MN3} or \cite{Review} for a more detailed explanation).
Taking, for instance, the sign of the coefficient of the KPZ non-linearity 
to be positive, the introduction 
of an upper wall leads to a (well established by now) set of critical exponents 
characterizing the, so called, multiplicative noise 1 (MN1) 
universality class \cite{MN1,MN2,MN3,MN4}.
On the other hand, a wall limiting negative values of the 
interface height (lower wall) leads
to a different type of phase transition as shown by T. Hwa and one of us
 some years back \cite{MN3}. 
In what follows, and following the nomenclature introduced in \cite{Review},
we will use the term MN2 to name this class.

It can be easily shown that a KPZ equation with positive non-linearity
and a lower wall is completely equivalent to a KPZ with
a negative non-linearity coefficient and an upper wall  \cite{Review}:
one just have to change the sign of the height variable in the KPZ-like equation
to verify this.

The MN2 class is of great importance in the context of alignment of DNA and other
biological sequences. It has been argued by Hwa and collaborators, that the 
phase transition appearing upon changing the, so called, scoring parameter in the 
commonly used alignment algorithm can be mapped into the MN2 
critical point \cite{Hwa}. It is also related to some instances 
of non-equilibrium wetting \cite{Haye1}. For some other applications
and physical instances within this class see \cite{Review} and
references therein.

While the MN1 class has been extensively studied, specially after its connections
with non-equilibrium wetting \cite{Haye1,Haye2,Lisboa}
and with the problem of synchronization
in extended systems were established \cite{synchro,Review},
the MN2 class remains poorly studied. 
Furthermore, recent numerical analysis have revealed that the
preliminary critical exponent values reported in \cite{Hwa} might be
far from their true asymptotic values.

Aimed at clarifying these issues
it is the purpose of this paper to analyze the MN2 phase transition 
in one dimensional systems using:
i)  mean-field and field-theoretical techniques and 
ii) numerical (Monte Carlo) analysis of different models claimed to belong 
to this class. 

Finally let us stress that if the wall becomes attractive, rather than simply 
bounding, new phenomenology might appear. This is particularly
interesting in the context of synchronization \cite{Review,synchro}. 
This possibility will also be discussed along the paper.

\section{The MN2 class}

Let us consider a KPZ equation with a positive non-linearity coefficient, 
$\lambda >0$, in the presence of a lower wall,
\begin{equation}
\partial_t h(x,t) = a + b ~e^{-ph} + D \nabla^2 h + \lambda (\nabla h)^2
+ \sigma \eta(x,t).
\label{BKPZ}
\end{equation}
where $h$ is a height variable, $a$ represents a constant drift 
while $ b ~e^{-ph}$ is a bounding  wall.
The parameter $p>0$ controls the wall penetrability, the limit 
$p \to \infty$ corresponding to a perfectly rigid (impenetrable) wall. 
It has been shown previously that for both the MN1 and the MN2 classes 
the magnitude of $p$ does not influence the asymptotic properties at 
criticality \cite{MN1,MN2} (its sign, however, is important, as it determines 
whether the wall is a lower or an upper one).
The same property applies to equilibrium systems ({\it i.e.}, 
for $\lambda =0$) \cite{lipowskyfisher}. 
$\eta$ is a stochastic white noise with 
$\langle \eta \rangle =0$ and 
$\langle \eta(x,t) \eta(x',t') \rangle = 2 \delta(t-t') \delta (x-x')$,
where $\langle \cdot \rangle$ denotes an average over the distribution
of the noise.  

For a fixed value of $b$ the interface experiences a pinning-depinning
transition at some value $a=a_c$. 

Some remarks concerning the connection of the previous equation 
with wetting problems follow.
When $h(x,t)$ is viewed as the distance separating a liquid-gas
interface from a solid wall, Eq. (\ref{BKPZ}) can then be
interpreted as a dynamic model for nonequilibrium wetting. Under this
perspective $a$ is the chemical potential difference between the 
liquid and the gas phases, $\langle h \rangle$ is the thickness of the 
wetting layer, and the wall is a rigid, physical substrate. 
At {\em bulk phase coexistence}, {\it i.e.} for the value of $a=a_c$ for 
which in the absence of the wall the interface does not move on average,
$\langle h \rangle$ diverges at all temperatures above certain 
wetting temperature, $T_W$, while for $a \not= a_c$ the thickness 
of the liquid film can be big, but finite (pinned interface). 
The temperature here is controlled by the parameter $b$, which
vanishes linearly with the mean-field wetting temperature as $T-T_W$. 
Thus, on approaching coexistence for $T>T_W$ 
($b>0$ at the mean-field level), $\langle h \rangle$
diverges as $\langle h \rangle \sim |a-a_c|^{\beta_h}$. 
This transition, termed {\it complete wetting}, 
is always continuous and the value of the $\beta_h$ exponent depends on 
the nature of the forces between the particles in the fluid phases 
and the wall. In this paper only short-range, exponentially decaying 
interactions between all the particles and the substrate (as described 
by Eq. (\ref{BKPZ})) are considered.
  
 The change of variables $n=\exp(-h)$ transforms Eq. (\ref{BKPZ}) into
the MN2 Langevin equation:
\begin{equation}
{\partial}_{t} n(x,t) =  
{\nabla}^{2} n - 2 {(\nabla n)^2 \over n}
-(a+1)n-bn^{p+1} +  n \eta,
\label{mnminus}
\end{equation}
where for the sake of simplicity we have 
set $\lambda=D=\sigma=1$ and Ito calculus \cite{Ito} has 
been used (different coefficients for the Laplacian and the KPZ non-linear 
term could be reabsorbed
using $n=\exp(-\alpha h)$ with a proper choice of $\alpha$). 
This transformation maps the depinning from the wall 
$\langle h \rangle \to \infty$ to a transition into an absorbing state
$\n \to 0$. 
The physical equivalence between the cases $\lambda>0$ with a 
lower-wall ($p>0$), and $\lambda<0$ with an upper wall ($p<0$),
reflects in the fact that the same equation is obtained 
using $n=\exp(-h)$ and $n=\exp h$, respectively.

Observe that Eq.(\ref{mnminus}) is identical to the MN equation
describing the MN1 class \cite{MN1,MN2,Review}, except for
the presence of an extra term  $ (\nabla n)^2 /n $.
This term can also be written as
$(\nabla n) \cdot (\nabla \ln(n)) = (\nabla n ) \cdot (\nabla h)$
suggesting that the interface language is the natural one for 
this class.
In fact, except for the factor 2 in front of $(\nabla n)^2 /n$
Eq. (\ref{mnminus}) coincides with the Cole-Hopf transform
of 
\begin{equation}
\partial_t h (x,t) = \nabla^2 h + a +b e^{-ph} + \eta(x,t)
\end{equation}
that describes the growth of wetting layers toward their 
equilibrium state \cite{lipowsky} 
(observe that this is just the equilibrium, Edwards-Wilkinson model, 
in the presence of a bounding wall). Note also that the factor $2$ 
in Eq. (\ref{mnminus}) cannot be readsorbed by reparametrizing.

Finally, let us underline that in the regime where $b<0$ 
the wall becomes attractive (which might be necessary to 
describe some physical situations as, for instance, synchronization problems
as said in the introduction) 
and a new term, say, $c \exp(-2h)$ (equivalently $c n^{2p+1}$) with $c>0$ 
has to be added to stabilize the equation. 

  Having presented the equations defining the model, in the forthcoming  sections
we study the associated physics by using i) mean field approaches, ii) 
field theory, and iii) numerical, Monte Carlo simulations 
combined with scaling arguments.

\section{Mean-field approaches}
Mean-field approaches to Eq. (\ref{BKPZ}) can be implemented with 
several degrees of sophistication. A crude approximation consists of 
ignoring the noise and spatial variations. At this level one trivially gets
that the order parameter $\langle h \rangle$ vanishes as $a \to 0$ with an
exponent $\beta_h=0$. When applied to Eq.(\ref{mnminus}), this
approximation yields a shifted critical point at $a_c=-1$ and the
usual result for the order parameter critical exponent, $\beta_n=1/p$.
But, as experience with other systems with multiplicative-noise dictates, 
neglecting completely the noise is a too crude approximation, that eliminates
most of the characteristic traits of MN physics.

Now the effect of allowing a spatially varying order
parameter and taking the noise into consideration 
are examined.
The Laplacian is discretized as
\begin{equation}
\nabla^2 n_i ={1 \over 2d} \sum_{j} (n_j-n_i) \approx \n 
-n_i,
\end{equation}
where the sum runs over the nearest neighbors of $i$ and a large 
system dimensionality has been assumed. Similarly, the square gradient term
can be written as
\begin{equation}
{(\nabla n)_i^2 \over n_i} \approx {\langle n^2 \rangle \over n_i}
-2\n + n_i.
\end{equation}
The one-site stationary probability distribution is then readily 
obtained from the associated Fokker-Planck equation,
\begin{eqnarray}
P_{st}\Big(n,\n\Big) &\propto& {1 \over n^2} \exp \int^n {
(\nabla x)_i^2-2(\nabla x)_i^2/ x_i -(a+1)x_i-bx_i^{p+1} \over x_i^2} \ dx_i,
\nonumber \\
&\approx &{1 \over n^{a+6}} \exp \Bigg(-5{\n \over n}-b {n^p\over p} 
+{\langle n^2 \rangle \over n^2}\Bigg).
\label{pmnminus}
\end{eqnarray}
where $\n$ and $\langle n^2 \rangle$ have to be calculated self-consistently.
For $a< a_c=-5$, $P_{st}$ is not normalizable what means that the 
stationary state is the absorbing phase $\n =0$. For 
$\n \not=0$, however, the non-analyticity of 
$\exp[\langle n^2 \rangle / n^2]$ at $n=0$ again renders $P_{st}$ 
non-normalizable. As a result, there is no well-defined active phase 
at this mean-field level for Eq. (\ref{pmnminus}). 
Similar problems are found if the same type of approach is 
applied to Eq. (\ref{BKPZ}) instead of Eq. (\ref{mnminus}).

To avoid the presence of the square gradient term in Eq. (\ref{mnminus}), 
and the complications of its, somewhat arbitrary, discretization, we resort to 
a different change of variables. After $n=\exp h$, 
\begin{eqnarray}
\partial_t n &=& \nabla^2 n +(a+1)n +bn^{3-p} +n \eta, \nonumber \\
P_{st} &\sim& {1 \over n^{2-a}} \exp\bigg[{b n^{2-p} \over 2-p}-{\n\over n}\bigg].
\label{mnplus}
\end{eqnarray}    
These equations are simpler than (\ref{mnminus}) and (\ref{pmnminus}),
but at the cost that $\n$ is no longer an order parameter
as, at the transition point $\n \to \infty$ rather than going to $0$.  
Furthermore, Landau expansions only make sense when
the order parameter vanishes at the critical point, making the whole approach
inconsistent.
One possible way to circumvent this problem is to monitor $m \equiv 1/n$, 
and study $\tilde{P}_{st}(m) dm =P_{st}(n) dn$,
but given the non-Gaussian nature of the probability distribution 
the substitution $\n = 1/\langle m \rangle$ is likely to be incorrect,
and there seem to be no safe way to proceed.

Summing up, non-trivial mean-field approaches detect some problems
with the model under consideration, and are not able to predict a phase
correct phase diagram. A sound mean-field approximation, needs therefore 
to be found. Notice that none of these problems occur in the case of a 
negative KPZ non-linearity, {\it i.e.} in the MN1 class, where a
standard mean-field approximation yields qualitatively 
correct results (see \cite{Review,Lisboa} and references therein).

\section{Field theoretical considerations}

 The Langevin equation for MN1 is known to be super-renormalizable,
{\it i.e.} Feynman diagrams can be computed to  all orders and resummated.
This does not imply that critical exponents can be computed in all the
cases, as the renormalization group flow-equation has runaway 
trajectories supposed to converge to a {\it strong  coupling  fixed point}. 
But at least, the correct phase diagram, including strong and weak coupling 
fixed points can be obtained. 

For the MN2 the situation is far more complicated, as can be {\it 
a priori} anticipated given the failure of mean-field approaches.
The extra term,  $ (\nabla n)^2 /n $, being singular in $n$ precludes 
the use of perturbative expansions around $n=0$. 
Given the lack of a non-perturbative approach to the KPZ and MN 
strong coupling fixed points, there is not much we can add to 
this section, except that there is a promising attempt to tackle this
and related KPZ-like problems.
There is a formalism, developed by Fogedby, 
aimed at developing a strong coupling theory for KPZ based 
on a semiclassical or WKB approximation applied upon the Martin-Siggia-Rose
generating functional \cite{fogedby}. Its main advantage is that it does not
involve expansions around classical noiseless solutions, but around
classical (extremal) noisy solutions. 
It would be very interesting to extend these ideas to KPZ problems 
in the presence walls,
namely, to the multiplicative noise universality classes MN1 and MN2.
 
\section{Numerical results}

Owing to the failure of standard (non-trivial) mean-field approaches and lacking
so far of an alternative analytical route, numerical methods are required to glean 
insight into the system properties. We have carried out simulations
of two surface growth models. Both of them, in the absence of walls, 
are known to belong to the KPZ universality class. 
An extra rule is then added to generate a bounding wall, as described above.

\subsection{Model 1}

Our first model was introduced in \cite{krugs} and is defined
as follows: the surface position at time $t$ above a site $x$ on a 
one-dimensional lattice of size $L$ is given by a continuous height 
variable $h_t(x)$. A new height configuration is then generated in 
a three-step process.
\begin{enumerate}
\item  Each lattice site $h_t(x)$ 
is updated according to $h'_t(x)=h_t(x)+a+\eta_t(x)$. $\eta_t(x)$ is
a random number uniformly distributed in [0,1] and $a$ is a constant drift
term analogous to that of (\ref{BKPZ}).
\item  The configuration is changed to  
$h_{t+1}(x)=\min[h'_t(x\pm 1)+\gamma,h'_t(x)]$, where $\gamma$ is
a constant whose precise value is not essential for the final results.
We have set without loss of generality $\gamma=0.1$  as in previous
numerical analyses.
\item  A hard wall at $h=0$ is introduced by way of the 
additional rule $h_t(x)=\min[h_t(x),0]$ \cite{MN3}.
\end{enumerate}
Finally, periodic boundary conditions are imposed and 
$h_t(x)$ is initially set to $0$.  

The continuum counterpart of this model is known to be a KPZ equation \cite{krugs}
 with $\lambda<0$ in the presence of an upper wall \cite{note} 
which, as remarked before, is
equivalent to the case $\lambda >0$ and a lower wall, and corresponds to 
the MN2 class. 
Numerical results for this model were first presented in \cite{MN3}
and seemed to be consistent with mean-field ({\it i.e.} single-site) like behavior.
However, a similar problem recently studied in the context of synchronization 
has revealed inconsistencies probably due to insufficient 
statistics \cite{Review,synchro}. 
In this subsection improved simulation results are provided upon revisiting the 
analysis reported in \cite{MN3} for larger system sizes and longer 
sampling times. 
The case of an attractive wall, not included in \cite{MN3}, 
is also considered.

First, we take up the case of a simple (non-attractive) wall, corresponding to
$b>0$ in Eqs. (\ref{BKPZ}) and (\ref{mnminus}).
Figure (1) shows how the steady order parameter $\n$ changes
with the system size $L$. Within the active phase, it saturates to
a constant value, while in the absorbing one it bends down and decays 
exponentially. Our best estimate for the critical point is 
$a_c=1.57433(2)$ and from
the slope of the curve we get $\beta_n/\nu_\bot \approx 0.33(2)$ \cite{note2}.
This value of $a_c$ corresponds to the point where a free interface 
(far from the wall) has zero average velocity.
The time evolution of $\n$ at the critical point for different system
sizes (Figure (2)) behaves like
$\langle n(t) \rangle \sim t^{-\theta_n}$, with $\theta_n =0.215(15)$ or,
equivalently, $\langle h(t) \rangle \sim t^{\theta_h}$, 
with $\theta_h =0.355(15)$.
As for the exponents $\beta$, which govern the saturation of the order
parameter within the active phase, they have been computed using
the largest available system sizes ($L=1600$ and $3200$). The best fit
to $ \langle X \rangle \sim |a-a_c|^{\pm \beta_X}$ yields
$\beta_n \approx 0.32 (3)$ (for $\langle n \rangle$) the and $\beta_h =0.52(2)$
(for $\langle h\rangle$), respectively.
The error margin is typically larger here than for other exponents due
to the sensitivity to the uncertainty in the determination of the 
critical point.

It was proved in \cite{MN1,MN2} that these exponents must satisfy 
the scaling relation $z= \beta/(\nu_\bot \theta)$, where
$z$ is the dynamic exponent of the KPZ equation.
The exact value for $z$ in $d=1$ is $3/2$, thereby   
$z = 0.33 / 0.215 = 1.5(1)$ in agreement with the prediction.
 In addition, also from \cite{MN1,MN2}, 
$\nu_\bot =1$, and from our direct measurements we get 
$(\beta_n/\nu_\bot)/ \beta_n = 0.33/0.32$, implying $\nu_\bot \approx 1$.
In terms of $h$ and assuming $\nu_\bot=1$, 
$z=0.52/.355= 1.5(1)$ which, again, is compatible with $3/2$ 
within error bars.

The two alternative, but equivalent, mathematical descriptions of the 
MN2 class in terms of $h$ and $n=\exp(-h)$ can be
related noting that the latter is essentially the 
density of sites at zero hight, $n(x,t)=\delta_{h(x,t),0}$ \cite{Haye3}.
We have verified that $n$ and $\delta_{h(x,t),0}$ 
exhibit the same asymptotic scaling behavior.

\begin{figure}
\vspace{0.5cm}
\includegraphics[width=8cm]{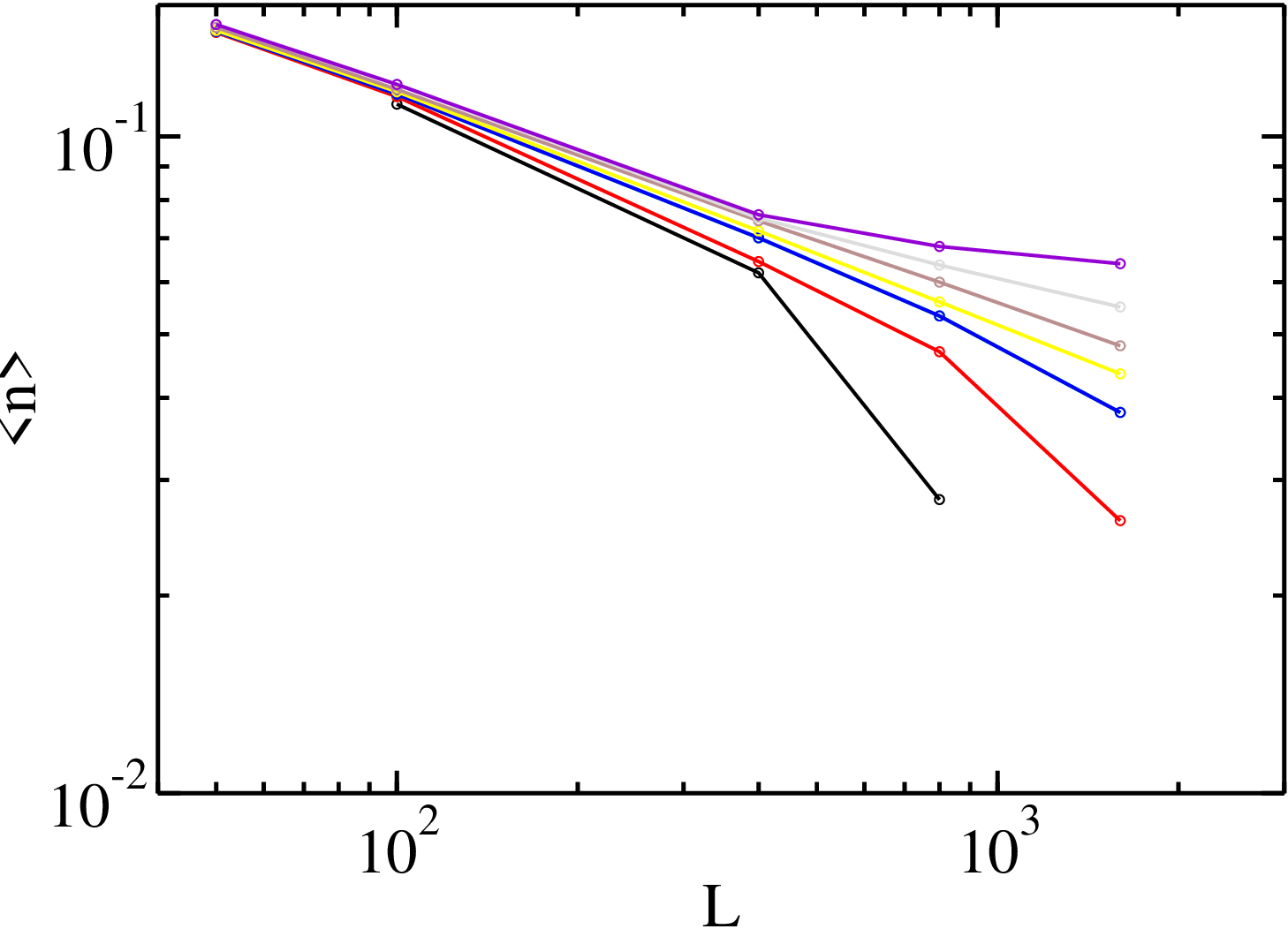}
\label{betanusin}
\caption{Model 1: steady-state values of the oder parameter, $\n =\langle
\exp h \rangle$, as function of the system size, $L$, for drifts 
(top to bottom) 1.57450, 1.57440, 1.57435, 1.57433,  
1.57430,  1.57425,  1.57420. $\langle \cdot \rangle$
denotes both spatial and temporal averages, as well as 
averages over independent runs. The straight line corresponds
to the critical point $a_c=1.57433(2)$ and from its slope
$\beta_n/\nu_\bot \approx 0.33(2)$.}
\end{figure}

\begin{figure}
\vspace{0.5cm}
\includegraphics[width=8cm]{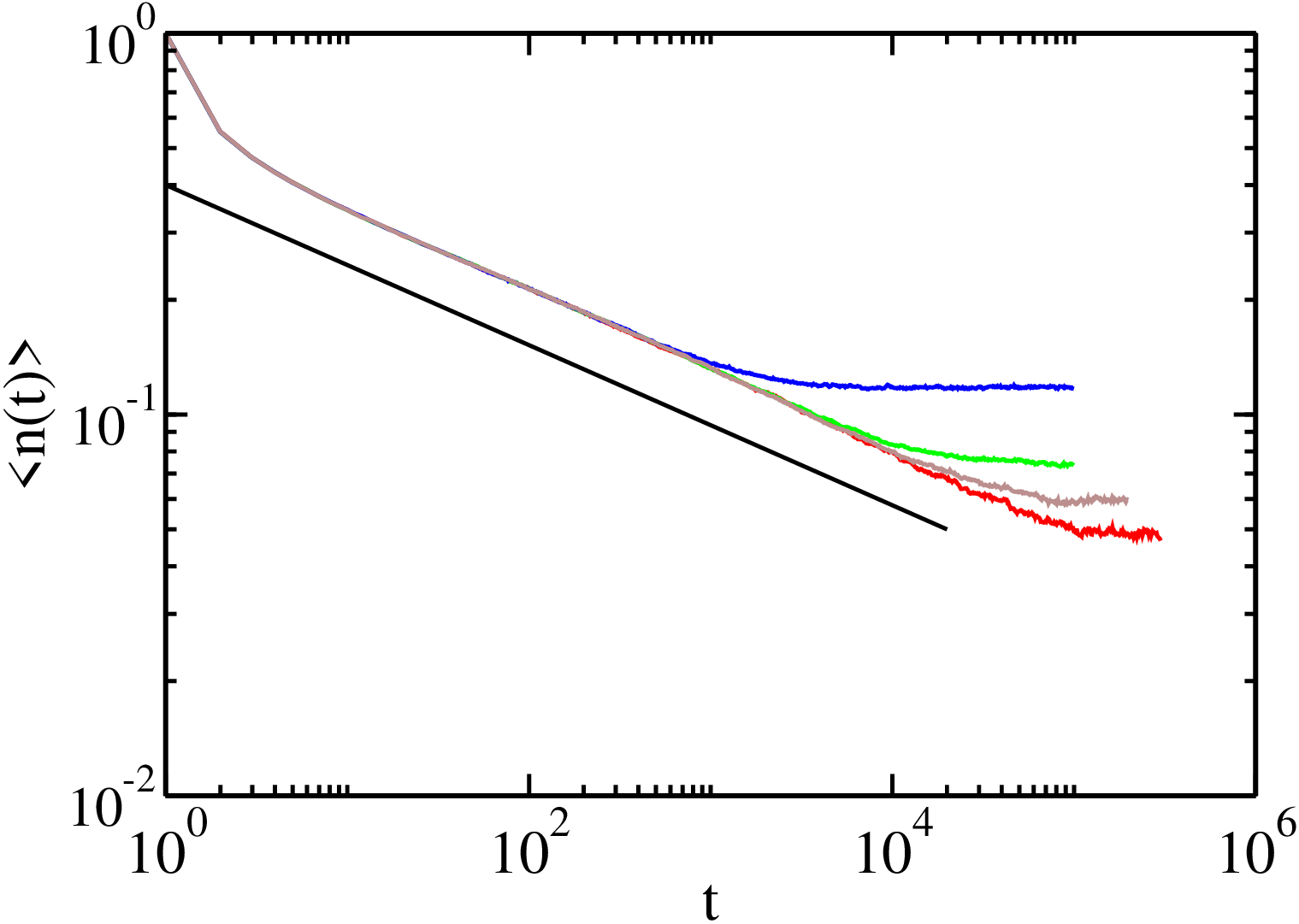}
\label{theta}
\caption{Model 1: time evolution of the order parameter, $\langle n(t) \rangle$,
at criticality, $a=a_c$, for system sizes 100, 400,
800, and 1600. The saturation values are those of Figure 1.
The best fit gives $\theta_n=0.215(15)$. The straight line
is a guide for the eye and has a slope -0.215.}
\end{figure}

\begin{figure}
\vspace{0.5cm}

\includegraphics[width=8cm]{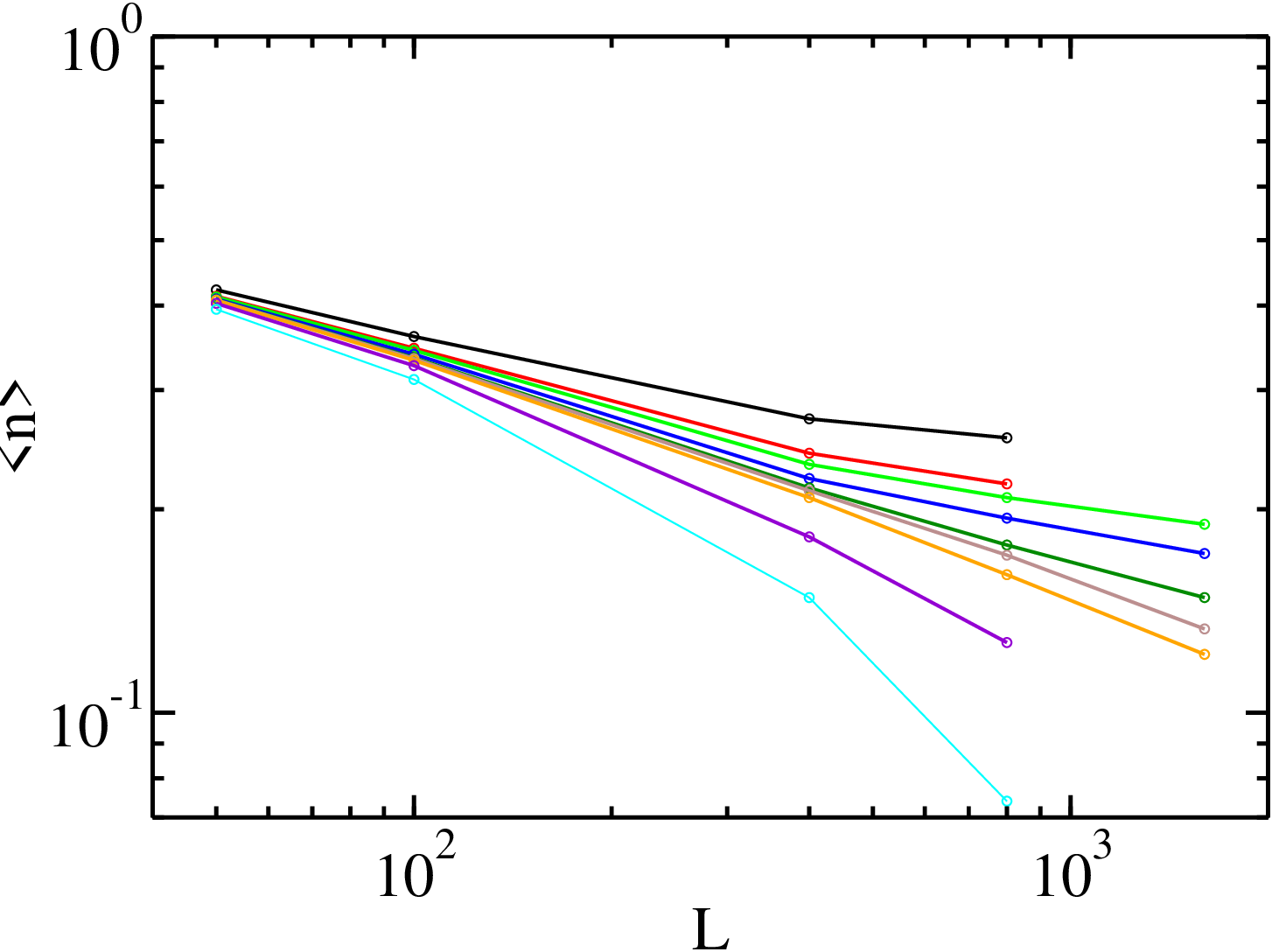}
\label{betanucon}
\caption{
Model 1: steady-state values of the oder parameter for an
attractive wall, $b=-0.3$, as function of the system size 
$L$ for drifts (top to bottom) 1.57450, 1.57447, 1.57445, 1.57440, 
1.57435, 1.57433, 1.57430, 1.57425, 1.57420. The best fit to a
straight line again corresponds to $a_c=1.57433$.}
\end{figure}

Attractive walls can also be simulated within this model 
by simply substituting $a$ by $a-b\delta_{h,0}$, 
where $b<0$ and the sign convention is 
chosen to keep the analogy with Eq. (\ref{BKPZ}). 
This means that wherever the interface in attached to the wall 
it experiences an additional (``sticky'') force pushing it against
the wall.
Extensive Monte Carlo simulations for $b=-0.3$ show
that the previous results, in what respect universal features, 
carry over without change, the only difference being that the approach to 
asymptotics is slower. Upon increasing the attractiveness of the wall
the transients become longer. 
The estimate for the critical point is the same one
as before (this is due to the fact that the free interface is not
affected by variations of the attractiveness parameter). Again, our best 
estimates for the critical exponents are $\beta_n/\nu_\bot =0.32$ 
and $\theta_n=0.215$ (see Figure (3)).
For $b=-0.4$ we still observe a second-order phase
transition with a crossover to the mentioned exponents.
For $b=-0.5$ transition becomes first-order but it still 
occurs at $a=a_c$. Within the active phase all the sites
are closed to the wall and the order parameter is $1$, but
it suddenly changes to $0$ upon decreasing $a$ and hysteresis
is observed for slightly subcritical values $a$ (the interface
is pinned for up to long times). Therefore, a tricritical point 
must exist between $0.4$ and $0.5$. We have identified 
it at $b=-0.42(1)$. 
The tricritical behavior has not been analyzed.


Let us stress that, contrarily to what happens for the MN1 class, where the presence
of an attractive wall induces a new and rich phenomenology (including
a broad region of phase coexistence and directed-percolation type of
transitions \cite{Review}), the addition of ``attractiveness'' has
a very mild effect here. Basically, it just shifts the position of 
the critical point
and induces a first-order phase transition for very strong attractions.

\begin{table}[t]
\begin{tabular}{lcccccc}
  &  \multicolumn{6}{c}{} \\
  \cline{2-7}
  & $\beta_n$ & $\nu_\bot$ & $\beta_n/\nu_\perp$ & $z$ & $\theta_n$
& $\eta$ \\
  \hline
 Model 1 ~~~ &$0.32(2)$& $0.97(5)$  & $0.34(2)$ & $1.55(5)$ &  $0.215(15)$
& not measured \\
 Model 2 ~~~ &$0.325(5)$& $\approx 1$  & $0.33(2)$ & $\approx 1.5$ &  $0.215(5)$
&$\eta = 0.8$
\\
 \hline
   \end{tabular}
\caption{Critical exponents for the  MN2 class in $d=1$.}
\end{table}

\subsection{Model 2}

In order to verify the robustness and eventual universality of the 
previous results, we have performed a second study of a different 
model. It is a restricted solid-on-solid (RSOS) model, 
a variant of the single-step model introduced in \cite{krugs}, in the 
presence of a wall.
 A similar model
has been recently studied in the context
of synchronization transitions \cite{Ahlers}, to study MN1 type of transitions.
Initially the wall is located at $h_w =0$ and a grooved interface
is placed beneath it, {\it i.e.} the interface has negative height  at all the 
positions.  

The dynamics proceeds as follows:
At each time step, a site is randomly picked from a one-dimensional lattice of
length $L$ and its height decreased two units, $h(i) \to h(i)-2$,
provided that $h(i)>h(i+1)$ and $h(i)>h(i-1)$, {\it i.e.} provided
that it is a local maximum. Should the
the RSOS constraint be violated, the trial is discarded and repeated. 
Every $2(L-1)/[1-2\delta v (1-L^{-1})]$ steps the wall retreats one
unit and, simultaneously, the interface is moved downwards by two units
wherever it lies above the wall \cite{Ahlers}. 
The difference between the wall and interface velocities, 
$\delta v$, acts as the control parameter: if $\delta v$ is negative, 
the interface eventually depins from the wall, while for 
$\delta v>0$ it remains pinned). It can be easily shown, using random walk 
arguments that for the chosen wall velocity the system seats at its
critical point. By varying it, we have a control parameter.  
The possibility of computing analytically the critical point largely simplifies
the numerical analysis, and makes of this an efficient
discrete model.

The quantities monitored are $\langle \exp(h_w-h) \rangle$
and $h_w-h$. We have measured the exponents $\beta_n$ and $\theta_n$
for a system of $L=2^{20}$ sites. Our results lead to 
$\beta_n=0.325(5)$ and $\theta_n=0.215(5)$, in excellent agreement 
with those of model 1 (Figure (4)). 
Once more, assuming $\nu_\bot=1$, we get $z \approx 1.51$ in good
agreement with the scaling laws.
In addition, we have also measured the spreading exponent, $\eta$, 
that characterizes the number of pinned sites. 
It is computed averaging over all the runs and starting with an 
initial condition with a single point attached to the wall.  
Our best estimate is $\eta =0.80(2)$ (Figure (5)).
Measuring the surviving probability, and therefore the exponents $\delta$ 
and $\zeta'$, is a delicate technical point because it is hard to
decide when the activity of a run has ceased. We have not tackled
this problem here.
Lastly, we obtain $\theta_h =0.34(5)$, which is again in good agreement
with the value reported for our Model 1.
\begin{figure}
\vspace{0.5cm}
\includegraphics[width=8cm]{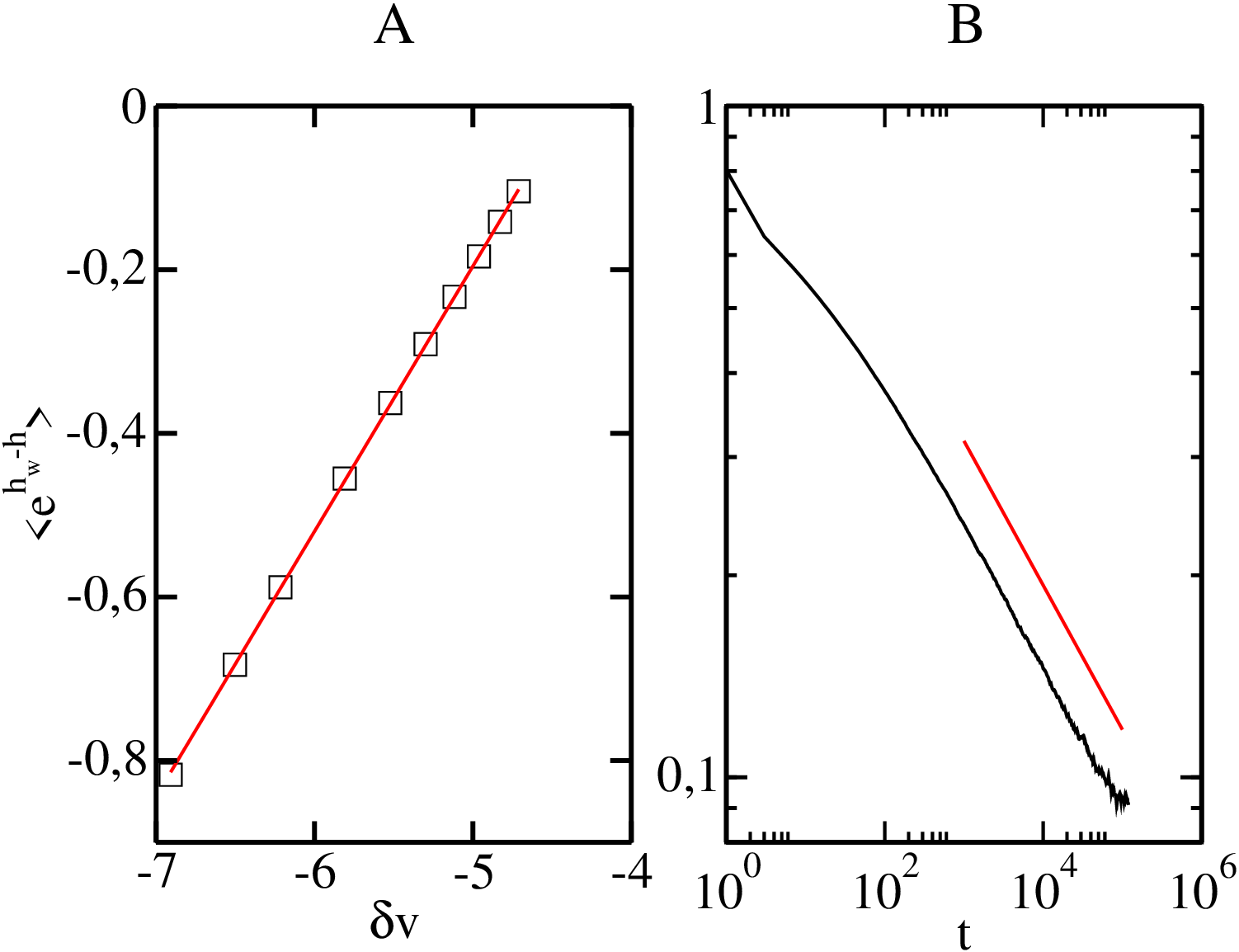}
\label{sidebyside}
\caption{
Model 2: (A) order parameter behavior, $\langle \exp(h_w-h) \rangle$, 
in the vicinity of the critical point $\delta v_c =0$. From 
the slope of the line we get $\beta_n =0.325(5)$.
(B) Decay of the order parameter, $\langle \exp(h_w-h) \rangle$,  
at the critical point yields $\theta_n=0.215(5)$ 
(cf. with Figure (2). The straight line is a guide for the
eye and has a slope $-0.215$.}
\end{figure}

\begin{figure}
\vspace{0.5cm}
\includegraphics[width=6cm]{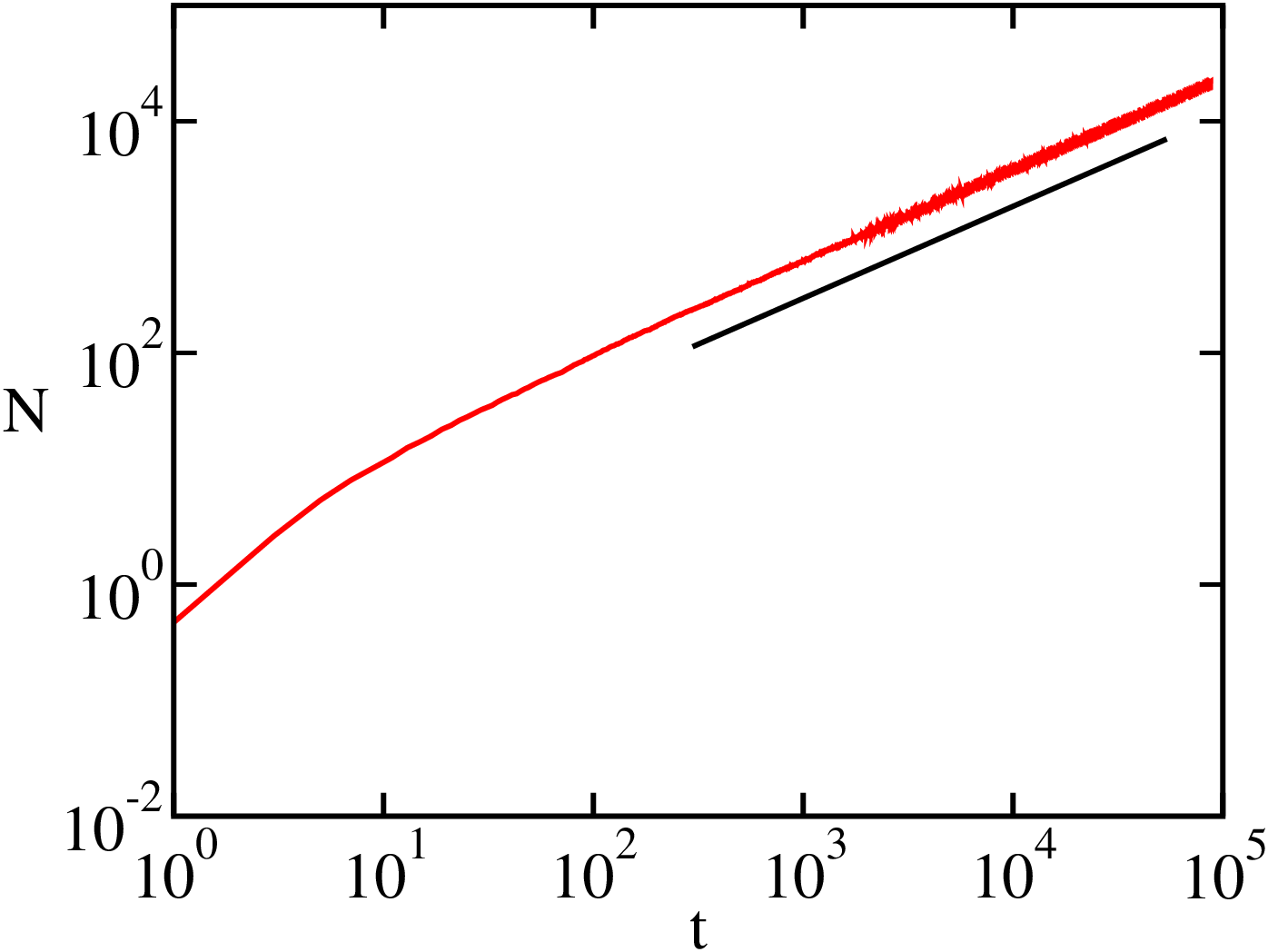}
\label{fattaheta}
\caption{
Model 2: number of pinned sites as a function of time for an
initial condition with a single point attached to the wall.   
A fit for late times yields $\eta=0.80(2)$. The straight line is a 
guide for the eye and has slope $0.8$.}
\end{figure}

 We have also considered different variations of the model in which the
interface can penetrate the wall at some points, {\it i.e.} the wall is not
perfectly rigid.  None of the universal features seem to be affected by this 
change.

In conclusion, 
all these results support strongly the existence of robust 
universality in the MN2 class.

As a matter of consistency, we have modified the algorithm of the
model to simulate a lower wall, and therefore a case expected to be
in the MN1 class. We obtain the set of exponents
$ \beta_n=1.69, \theta_n= 1.19$ and $\eta = -0.4$, all of them 
in agreement we previously reported results and showing
that the upper and lower problem belong to different universality
classes \cite{Review,MN2,MN4}.
 
\subsection{Numeric integration of stochastic differential equations}

As a further test for universality, 
we have numerically integrated Eq. (\ref{BKPZ}) using
a Milshtein's algorithm \cite{Maxi}. A system size of $L=2000$ was considered
and the time step and mesh size were set to $\Delta t =0.001$
and $\Delta x=1$, respectively. For simulation times up to $t=10^{6}$ 
($10^9$ trials per site) our results lie far from the asymptotic regime.
We do not discard numerical instabilities in the integration scheme,
as it is known that results from numerical integrations 
may not agree with the predictions from the continuum KPZ \cite{KPZproblem}.
The Cole-Hopf transform is a standard way to account numerically 
for the integration of bounded KPZ  equations and, indeed, Langevin equations
with MN are by far more stable than their interface-language KPZ-like counterparts
\cite{MN2,MN4}. Nevertheless, we have also found
numerical problems when integrating (\ref{mnminus}), either when the
extra term is written in logarithmic form or averaging for smoothing 
the gradient. All numerical attempts are unstable nearby the absorbing state, 
owing to the presence of the extra singular term. We leave, therefore,
the numerical integration of a continuous Langevin equation, representative  
of the MN2 class as an open, challenging problem.

\section{Discussion}

We have characterized the MN2 universality class, or analogously its 
bounded KPZ counterpart,
which, as commented above, accommodates different physical phenomena.
We have studied it from, somehow deceptive mean-field and field theoretical
approaches, as well as by numerical studies. 
None of the analytical methods provides a satisfactory description
of the phase transition present in this class.
On the other hand, Monte Carlo  simulations of two different discrete 
interface models, argued to  belong to   this universality class, 
give a firm evidence for the existence of a robust universality class.
Contrarily to what previous simulations seemed to indicate \cite{MN3}, our 
results are not a simple extension of the ones obtained for one-site, 
implying that spatial correlations play an important role.
Table 1 gathers the values of the critical exponents in terms of 
$n$ for the two discrete models considered in this paper. 
From the Monte Carlo estimates, 
it cannot be discarded that they adopt the rational values $\beta_n=1/3$ and 
$\theta_n=2/9$, which combined with  the exact values 
derived in \cite{MN1,MN2},
$\nu_\bot=1$ and $z=3/2$, would lead to  $\beta_n/\nu_\bot= 1/3$.
Note that our results do not compare well with those of 
the nonequilibrium wetting model reported in \cite{Haye2}. 
We believe this is probably due to the extremely long transients known
to be present in that model. 
We have verified that the transition 
point is located at the same value of the control parameter 
for any value of the ``attractiveness'' parameter ($b$ in Model 1), either representing
an attractive or a non attractive wall.
For strong enough attractive walls, {\it i.e.}
$b$ sufficiently negative, the transition becomes first order as 
in \cite{Haye2}, while if the wall is weakly attractive then it
remains in the MN2 class. 

The problem of reaching  a satisfactory analytical understanding,
and even that of obtaining sound results from numerical integrations 
of the continuous Langevin-equation (in either the interface or 
the density language) representative of this class remains
an open challenge. 

Summing  up, even though strong evidence is provided confirming 
the existence of a universality class ({\it i.e. } the corresponding critical
exponents are computed with good precision in one dimension and they are 
universal in two different discrete models), 
its theoretical description in terms of Langevin equations,
contrarily  to what happens for the closely related MN1 class,
is far from satisfactory. In particular, the Langevin equation does not seem 
to admit sound mean-field solutions, nor is amenable to be treated by means
of standard perturbative field theoretical tools, 
nor it admits a stable numerical integration. 
Identifying 
the physical causes at the root of these difficulties is a challenge
for future research.

{\centerline {\bf ACKNOWLEDGMENTS}}

\vspace{0.5cm}
F.S. acknowledges financial support from the Funda\c c\~ao para a 
Ci\^encia e a Tecnologia, contract SFRH/BPD/5654/2001.
Financial support from the Spanish MCyT (FEDER) under project BFM2001-2841,
and from the AECI,
are also acknowledged.


\begin{thebibliography}{}



\bibitem{Sancho}
See J. Garc{\'\i}a-Ojalvo, and J. M. Sancho, 
{\it Noise in Spatially Extended Systems},
Springer, New York, 1999; and references  therein. See also, J. M. Sancho
and J. Garc{\'\i}a-Ojalvo, in Lecture Notes in Physics {\bf 557}, p.235, ed.
J. A. Freund  and T. P\"oschel, Springer-Verlag, Berlin (2000).

\bibitem{Review}
M.A. Mu\~noz, preprint 2003, cond-mat/0303650.

\bibitem{KPZ}
M. Kardar, G. Parisi and Y. C. Zhang,
Phys. Rev. Lett. {\bf 56}, 889 (1986).

\bibitem{HZ}
T. Halpin-Healy and Y.-C. Zhang, Phys. Rep. {\bf 254}, 215 (1995); and 
references therein.

\bibitem{Barabasi}
A. L. Barab\'{a}si,  H. E. Stanley,
{\it Fractal Concepts in Surface Growth}
Cambridge University Press, Cambridge, 1995; and references therein.

\bibitem{MN1} G. Grinstein, M.A. Mu\~noz, and Y. Tu
Phys. Rev. Lett. {\bf 76}, 4376 (1996).

\bibitem{MN2} Y. Tu, G. Grinstein and M.A. Mu\~noz,
Phys. Rev. Lett. {\bf 78}, 274 (1997).

\bibitem{MN3} M.A. Mu\~noz and T. Hwa,
Europhys. Lett. {\bf 41}, 147 (1998).

\bibitem{MN4}
W. Genovese and M.A. Mu\~noz,
Phys. Rev. E {\bf 60}, 69 (1999).


\bibitem{Haye1}
H. Hinrichsen, R. Livi, D. Mukamel, and A. Politi,
Phys. Rev. Lett. {\bf 79}, 2710 (1997).

\bibitem{Haye2}
H. Hinrichsen, R. Livi, D. Mukamel, and A. Politi,
Phys. Rev. E {\bf 61}, R1032 (2000).

\bibitem{Lisboa}
F. de los Santos, M.M. Telo da Gama, and M.A. Mu\~noz,
Europhys. Lett. {\bf 57}, 803 (2002);
 Phys. Rev. E {\bf 67}, 021607 (2003);
Proceedings of the 7th Granada Seminar on Computational Physics.
Ed. J. Marro and P. L. Garrido; Am. Inst. of Phys. 661 (2003).
Cond-mat/0211124.

\bibitem{synchro} M.A. Mu{\~n}oz and R. Pastor Satorras,
Preprint. Cond-mat/0301059.

\bibitem{Hwa} T. Hwa and M. Lassig, Phys. Rev. Lett. {\bf 76}, 2591 (1996).
See also, T. Hwa and M. Lassig, ``Optimal Detection of Sequence Similarity by Local Alignment"
in Proceedings of the Second Annual Int. Conf. on Computational
Molecular Biology (RECOMB98),
S. Istrail, P. Pevzner, and M.S. Waterman eds, 109-116 (ACM Press, 1998); and
references therein.
R. Olsen, T. Hwa and M. Lassig,
"Optimizing Smith-Waterman Alignments"
in Pacific Symposium on Biocomputing 4, 302-313 (1999).

\bibitem{lipowskyfisher}
R. Lipowsky and M.E. Fisher, Phys. Rev. B {\bf 36}, 2126 (1987).

\bibitem{Ito} The sole difference between utilizing the Ito or the 
Stratonovich conventions, in this case, is a trivial shift in 
$a$ \cite{VK,Gardiner}. 

\bibitem{VK} N.G. van Kampen, {\it Stochastic Processes in Physics
and Chemistry}, North Holland, Amsterdam, 1981.

\bibitem{Gardiner} C.W. Gardiner,
{\it Handbook of Stochastic Methods}, Springer Verlag,
Berlin and Heidelberg, 1985.

\bibitem{lipowsky}
R. Lipowsky, J. Phys. A {\bf 18}, L585 (1985).

\bibitem{fogedby} H. D. Fogedby, Phys. Rev. E.  {\bf 57}, 4943 (1998). 
 H. D. Fogedby, Cond-mat/0303632.

\bibitem{krugs} 
J. Krug, Adv. in Phys. {\bf 46}, 139 (1997).
J. Krug and H. Spohn, in {\it Solids far from equilibrium}, Ed. C. Godr\`eche,
Cambridge University Press, (1991).

\bibitem{note} Observe that the presence of the function ``$\min$'' is the second
step of the algorithm induces a negative average velocity, implying  that in
the continuum counterpart $\lambda$  has to be negative. On the other
hand the last step $\min{h(x,t),0}$ obviously generates an upper wall.

\bibitem{note2} We use the subscript $_n$ to denote exponents related to the order
parameter $n$ and $_h$ for exponents associated with the average height.

\bibitem{Haye3}
H. Hinrichsen, cond-mat/0302381.

\bibitem{Ahlers} V. Ahlers and A. Pikovsky, Phys. Rev. Lett.
{\bf 88}, 254101 (2002). V. Ahlers, Ph. D. thesis.
http:// www.stat.physik.uni-potsdam.de/~volker/publ.html.
F. Ginelli, {\it et al.}, cond-mat/0302588.

\bibitem{Maxi} M. San Miguel and R. Toral, {\it Stochastic Effects in
Physical Systems}, to be published in {\it Instabilities  and
Nonequilibrium Structures}, VI, E. Tirapegui and W. Zeller,
eds. Kluwer Academic Pub. (1997). (Cond-mat/9707147).

\bibitem{KPZproblem} T. J. Newman and A. J. Bray,
J. Phys. A {\bf 29}, 7917 (1996). C.H. Lam and F. G. Shin,
Phys. Rev. E {\bf 58}, 5592 (1998).

\end{thebibliography}
\end{document}